\documentstyle[12pt]{article}

\begin{document}
\title{Cosmic String in Scalar-Tensor Gravities}
\author{Maria Em\'{\i}lia X. Guimar\~aes \\
\mbox{\small{Laboratoire de Gravitation et Cosmologie Relativistes}} \\
\mbox{\small{Universit\'e Pierre et Marie Curie - CNRS/URA 769}} \\
\mbox{\small{Tour 22/12, 4\`eme \'etage, BC 142}} \\
\mbox{\small{4, Place Jussieu 75252 Paris cedex 05, France}}}
\date{}
\maketitle
\begin{abstract}
We consider a local cosmic string described by the Abelian-Higgs model in the framework of scalar-tensor gravities. We find the metric of the cosmic string 
in the weak-field approximation. The propagation of particles and light is analysed in this background. 
This analysis shows that the (unperturbed) cosmic string 
in scalar-tensor theory presents some analogous features to the wiggly cosmic 
string in General Relativity. 

{\em PACS number: 0450, 9880 C}
\end{abstract}

\section{Introduction}

The scalar-tensor theories of gravity proposed by Bergmann \cite{be}, Wagoner 
\cite{wa} and Nordtverdt \cite{no} - generalizing the original Brans-Dicke 
\cite{bra} theory - have been considerably revived in the last years. Indeed, 
the existence of a scalar field as a spin-0 component of the gravitational 
interaction seems to be a quite natural prediction of unification models such 
as supergravity or superstrings \cite{sup}. Appart from the fact that 
scalar-tensor theories may provide a solution for the problem of terminating inflation \cite{la,ca}, these theories by themselves have direct implications 
for cosmology and for experimental tests of the gravitational interaction: One 
expects that in the Early Universe the coupling to matter of the scalar 
component of the gravitational interaction would be of the same order of the 
coupling to matter of the long-range tensor component although in the present 
time the observable total coupling strength of scalars $(\alpha^2)$ is 
generically small \cite{da}. Besides, any gravitational phenomena will be 
affected by the variation of the gravitational ``constant" 
$G_{eff} \sim \tilde{\phi}^{-1}$. So, it seems worthwhile to analyse the 
behaviour of matters in the framework of scalar-tensor theories, specially 
those which originated in the early Universe such as topological defects. In 
this context, some authors \cite{gun} have considered the solutions of 
cosmic string and domain walls in Brans-Dicke theory.

The aim of this paper is to study the modifications on the metric of a local 
cosmic string in more general scalar-tensor gravities. These modifications are 
induced by the coupling of a scalar field to the tensor field in the 
gravitational lagrangean. For simplicity, we will consider a class of 
scalar-tensor theories where the potential $V(\tilde{\phi})$ (or as in 
Wagoner's notation, the potential $\lambda(\tilde{\phi})$ \cite{wa}) is vanishing. 

The action describing these theories is (in Jordan-Fierz frame) 
\begin{equation}
{\cal S} = \frac{1}{16\pi} \int d^4 x {\sqrt{\tilde{g}}} \left[ \tilde{\phi}
\tilde{R} - \frac{\omega(\tilde{\phi})}{\tilde{\phi}} \tilde{g}_{\mu\nu}
\partial_{\mu}\tilde{\phi}\partial_{\nu}\tilde{\phi} \right] 
+ {\cal S}_{m}[\Psi_m, \tilde{g}_{\mu\nu}] ,
\end{equation}
where $\tilde{g}_{\mu\nu}$ is the physical metric in this frame, $\tilde{R}$ 
is the curvature scalar associated to it and ${\cal S}_{m}$ denotes the 
action of the general matter fields $\Psi_m$. These theories are metric, which 
means that matter couples minimally to $\tilde{g}_{\mu\nu}$ and not to 
$\tilde{\phi}$. For many reasons it is more convenient to work in the 
so-called Einstein (conformal) frame, in which the kinematic terms of tensor 
and scalar fields do not mix  
\begin{equation}
{\cal S} = \frac{1}{16\pi G}\int d^4 x \sqrt{g}
[R - 2g^{\mu\nu}\partial_{\mu}\phi\partial_{\nu}\phi] + {\cal S}_m
[\Psi_m, A^2(\phi)g_{\mu\nu}]
\end{equation}
where $g_{\mu\nu}$ is the (unphysical) metric tensor in Einstein frame, $R$ is 
the curvature scalar associated to it and $A(\phi)$ is an arbitrary function 
of the scalar field. Action (2) is obtained from (1) by a conformal 
transformation in the physical metric  
\begin{equation}
\tilde{g}_{\mu\nu} = A^2(\phi) g_{\mu\nu}
\end{equation}  
and by a redefinition of the quantities 
\[
GA^2(\phi) = \tilde{\phi}^{-1} , \]
\[
\alpha^2 \equiv \left ( \frac{\partial{\ln A(\phi)}}{\partial\phi} 
\right )^2 = [2\omega(\tilde{\phi}) +3]^{-1} .
\]
It is important to remark that $\alpha(\phi)$ is the field-dependent coupling 
strength between matter and scalar fields. In the particular case of 
Brans-Dicke theory, $A(\phi)$ has the following dependence on 
$\phi \,\, :  \,\, A(\phi) = e^{2\alpha\phi}$, with
$\alpha(\phi)=\alpha= [2\omega +3]^{-1/2}$
=constant. In the Einstein 
frame, the field equations are written as the following  
\begin{eqnarray}
R_{\mu\nu} & =  & 2\partial_{\mu}\phi\partial_{\nu}\phi +  
8\pi G(T_{\mu\nu} -\frac{1}{2}g_{\mu\nu}T) , \nonumber \\
\Box_g \phi & = & -4\pi G\alpha(\phi)T .
\end{eqnarray}
We note that the first of the above equations can also be written in terms of the Einstein 
tensor $G_{\mu\nu}$ 
\[
G_{\mu\nu}=2\partial_{\mu}\phi\partial_{\nu}\phi- g_{\mu\nu}g^{\alpha\beta}
\partial_{\alpha}\phi\partial_{\beta}\phi + 8\pi GT_{\mu\nu} .
\]
The energy-momentum tensor is defined as usual
\[
T_{\mu\nu} \equiv \frac{2}{\sqrt{g}} \frac{\delta {\cal S}_m [A^2(\phi)
g_{\mu\nu}]}{\delta g^{\mu\nu}} ,
\]
but in the Einstein frame it is no longer conserved 
$\nabla_{\nu}T^{\nu}_{\mu} = \alpha(\phi)T\nabla_{\mu}\phi$. 
It is clear from transformation (3) that we can related quantities from both 
frames such that $\tilde{T}^{\mu\nu}=A^{-6}T^{\mu\nu}$ and  
$\tilde{T}^{\mu}_{\nu}= A^{-4}T^{\mu}_{\nu}$. 

In what follows we will search for a regular solution of an isolated static 
straight cosmic string in the scalar-tensor gravity described above. Hence, 
the cosmic string arises from the action of the Abelian-Higgs model where a 
charged scalar Higgs field $\Phi$ minimally couples to the $U(1)$ gauge 
field $A_{\mu}$
\begin{equation}
{\cal S}_{m} = \int d^4 x \sqrt{\tilde{g}} \left[ \frac{1}{2}D_{\mu}\Phi D^{\mu}
\Phi^* - \frac{1}{4}F_{\mu\nu}F^{\mu\nu}-V(\mid \Phi\mid) \right] ,
\end{equation}
with $D_{\mu} \equiv \partial +ieA_{\mu} , \,\, F_{\mu\nu} \equiv \partial_{\mu}
A_{\nu} -\partial_{\nu}A_{\mu}$ and the Higgs potential $V(\mid\Phi\mid) =
\lambda(\mid\Phi\mid^2 -\eta^2)^2$. $e,\lambda$ and $\eta$ are positive 
constants, $\eta$ being the characteristic energy scale of the symmetry 
breaking (eg, for typical GUT strings, $\eta \sim 10^{16} GeV$). 

We confine our attention to the static configurations of vortex type about 
the $z$-axis. In cylindrical coordinate system $(t,z,\rho,\varphi)$ such that  
$\rho \geq 0$ and $0 \leq \varphi <2\pi$, we impose the following form for 
the Higgs $\Phi$ and the gauge $A_{\mu}$ fields \footnote{With winding 
number $n=1$.} 
\begin{equation}
\Phi \equiv R(\phi)e^{i\varphi} \;\;\; and \;\;\; 
A_{\mu} \equiv \frac{1}{e}[P(\rho) -1]\delta^{\varphi}_{\mu} , 
\end{equation}
where $R,P$ are functions of $\rho$ only. Moreover, we require that these functions are regular everywhere and that they satisfy the usual boundary conditions for vortex solutions \cite{ni}
\[
R(0) = 0 \;\;\;\; and  \;\;\;\; P(0)=1 ,
\]
\begin{equation}
\lim_{\rho\rightarrow\infty} R(\rho) =\eta \;\;\; and 
\;\;\; \lim_{\rho\rightarrow\infty} P(\rho) =0. 
\end{equation}

In General Relativity, a metric for a cosmic string described by the 
action (5) above has been already found in the asymptotic limit by Garfinkle 
\cite{ga} and exactly by Linet \cite{li} provided the particular relation 
$e^2=8\lambda^2$ between the constants $e$ and $\lambda$ is satisfied. The 
question is whether one can find a solution for the cosmic string in the 
framework of the scalar-tensor gravity. The answer for this question seems 
obviously negative. However, we can consider the weak-field approximation for 
this solution in the same way as Vilenkin \cite{vi} in the framework of 
General Relativity. In fact, the weak-field approximation breaks down at 
large distances from the cosmic string. Therefore, we assume that at large distances the $\phi$ dependence on the right-hand side of the first of 
Einstein eqs. (4) must dominate over the $T^{\mu}_{\nu}$ term 
\footnote{As shown by Laguna-Castillo and Matzner \cite{lag}, the 
$T^{\mu}_{\nu}$ components vanish far from the cosmic string in General 
Relativity. Gundlach and Ortiz \cite{gun} showed that this feature remains 
valid for a cosmic string in Brans-Dicke theory. One must expect that this 
assumption can be applied to general scalar-tensor theories.}. So, we will 
neglect the energy-momentum tensor in Einstein eqs. (4) and find the vacuum 
solution as an asymptotic behaviour of $g_{\mu\nu}$ and $\phi$ and, then, 
match this vacuum metric with the metric of the cosmic string in the 
weak-field approximation. 

The plan of this work is as follows. In section 2, we find the exact vaccum 
metric for the Einstein eqs. (4). In section 3, we find the metric of the 
cosmic string in the weak-field approximation and analyse under which 
conditions it can be matched to the vacuum metric. We then analyse the propagation of particles and light in the linearized metric of the cosmic string. Our main 
result is that particles and light propagate in this background in a similar 
way as they propagate in the background of a wiggly cosmic string in 
General Relativity.

\section{The vacuum metric in scalar-tensor: an exact solution}

In this section we will find the exact static vacuum metric which is 
solution to the Einstein eqs. (4) in scalar-tensor theories. This metric is supposed to 
match the weak-field solution of the cosmic string, so it seems natural to 
impose the same symmetries to this vaccum spacetime as those of the string. 
We write the following metric 
\begin{equation}
ds^2 = g_1(\rho) dt^2 -g_2 (\rho)dz^2-d\rho^2 - g_3 (\rho)d\varphi^2 , 
\end{equation}
where $g_1 , g_2 ,g_3$ are functions of $\rho$ only, and $(t,z,\rho,\varphi)$ 
are cylindrical coordinates such that 
$\rho \geq 0$ and $0 \leq \varphi <2\pi$. 
Defining $u \equiv (g_1g_2g_3)^{1/2}$, the Einstein eqs. (4) in vacuum can 
be written as 

\begin{equation}
R^i_i = \frac{1}{2u} \left[ u\frac{g'_i}{g_i}\right]' =0 , \;\; (i=t,z,\varphi)
\end{equation}
\begin{equation}
G^{\rho}_{\rho} = -\frac{1}{4}\left[ \frac{g'_1g'_2}{g_1g_2} +
\frac{g_1'g_3'}{g_1g_3} + \frac{g_2'g_3'}{g_2g_3} \right]= -(\phi')^2 , 
\end{equation}
\begin{equation}
\frac{1}{u}\frac{d}{d\rho}(u\phi')=0 ,
\end{equation}
where the prime means derivative with respect to $\rho$. Besides, since 
$T^{\mu}_{\nu}=0$, the following expression is also valid
\begin{equation}
\sum_{i}R^i_i = \frac{u^{''}}{u} = 0 \;\;\; (i=t,z,\varphi).
\end{equation}
>From (12) it follows that $u$ is a linear function of $\rho \;\; (u \sim 
B\rho)$. This result enables us to solve eqs. (9) and (11) to find
\begin{equation}
g_i = k_i^{(0)} \left(\frac{\rho}{\rho_0} \right)^{k_i} \;\;\; and \;\;\;  
\phi =\phi_0 +\kappa \ln (\rho/\rho_0) ,
\end{equation}
respectively. $B, k_i^{(0)}, k_i$ and $\kappa$ are constants to be 
determined later. Combining the solution for $u$ with solutions (13),  
we obtain the following relations between the constants
\[
[k_1^{(0)}k_2^{(0)}k_3^{(0)}]^{1/2}=B ,
\]
\[
k_1 +k_2 +k_3 = 2,  \;\; and 
\]
\[
k_1k_2 +k_1k_3 +k_2k_3 = 4\kappa^2.
\]
Moreover, if we suppose the boost invariance of the string along the $z$-axis 
(i.e., $g_1=g_2$) the above relations are simplified and we finally find 
\[
k_1^{(0)}[k_3^{(0)}]^{1/2} =B,
\]
\[
k_3=2-2k_1 \;\; and 
\]
\begin{equation}
\kappa^2 =k_1(1-\frac{3}{4}k_1).
\end{equation}
The constant $k_1^{(0)}$ can always be absorbed by a redefinition of $t$ and 
$z$. Then, we obtain the final form for the vacuum metric
\begin{equation}
ds^2= \left(\frac{\rho}{\rho_0}\right)^{k_1}(dt^2-dz^2)-d\rho^2-
\left(\frac{\rho}{\rho_0}\right)^{2-2k_1}B^2d\varphi^2 ,
\end{equation}
in which $k_1$ (and consequently $k_3$), $B$ and $\kappa$ will be fully 
determined after the introduction of matter fields. 

The Ricci tensor in the vacuum metric (15) is regular (vanishes 
everywhere) if and only if $\phi=\phi_0$= const. (i.e, $\kappa=0$). This 
should not be surprising in views of the structure of Einstein eqs. (9-11), 
in particular eq. (10). Besides, $\kappa =0$ implies that the only allowed 
values for $k_1$ are $k_1=4/3$ and $k_1=0$. This latter value corresponds to 
the conical metric and only in this case $B^2=k_3^{(0)}$ can be interpreted as 
the deficit angle. As we will see in the next section, the values $k_1=4/3$ and 
$k_1=0$ are precisely the necessary conditions required to match metric (15) to 
the metric of the cosmic string in the weak-field approximation. Finally, we 
note that metric (15) can also be written in Taub-Kasner form \cite{ru} after 
a suitable coordinate transformation.

\section{Cosmic string solution in scalar-tensor gravity: the weak-field 
approximation}

We start by writting down the full Einsteins equations
\begin{equation}
R^t_t=R^z_z=\frac{1}{2u} \left[ u\frac{g_1'}{g_1}\right]' =8\pi G 
(T^t_t -\frac{1}{2}T) ,
\end{equation}
\begin{eqnarray}
R^{\rho}_{\rho} & = & \frac{1}{2} \left[ 2 \left(\frac{g_1^{''}}{g_1}
\right) -\left(\frac{g_1'}{g_1}\right)^2 +\left(\frac{g_3^{''}}{g_3}\right)
-\frac{1}{2}\left(\frac{g'_3}{g_3}\right)^2\right] \nonumber \\
& & = -2 (\phi')^2 + 8\pi G (T^{\rho}_{\rho}-\frac{1}{2}T) ,
\end{eqnarray}
\begin{equation}
R^{\varphi}_{\varphi}= \frac{1}{2u}\left[ u\frac{g'_3}{g_3}\right]' =
8\pi G (T^{\varphi}_{\varphi} -\frac{1}{2}T) ,
\end{equation}
\begin{equation}
\frac{1}{u}[u\phi']'=4\pi G\alpha(\phi)T.
\end{equation}
where the non-vanishing components of the energy-momentum tensor are
\[
T^t_t=T^z_z=\frac{1}{2}A^2(\phi) \left[(R')^2 +g_3^{-1}R^2P^2+\frac{1}{e^2}
g_3^{-1}A^{-2}(P')^2+2\lambda A^2(R^2-\eta^2)^2\right] ,
\]
\[
T^{\rho}_{\rho}= -\frac{1}{2}A^2(\phi)\left[ (R')^2-g_3^{-1}R^2P^2+
\frac{1}{e^2}
g_3^{-1}A^{-2}(P')^2-2\lambda A^2(R^2-\eta^2)^2\right] ,
\]
\begin{equation}
T^{\varphi}_{\varphi}= \frac{1}{2}A^2(\phi)\left[(R')^2- g_3^{-1}R^2P^2 -
\frac{1}{e^2}g_3^{-1}A^{-2}(P')^2+2\lambda A^2(R^2-\eta^2)^2 \right].
\end{equation}

The matter fields equations can be written as (where form (6) for a vortex 
type solution is imposed)
\begin{equation}
R^{''}+R' \left[ \frac{g_1'}{g_1}+\frac{1}{2}\frac{g_3'}{g_3}-2\alpha(\phi)
\phi' \right] - R [g^{-1}_3 P^2 +4\lambda A^2 (R^2-\eta^2)] =0 ,
\end{equation}
\begin{equation}
P^{''} +P' \left[ \frac{g_1'}{g_1} +\frac{1}{2}\frac{g_3'}{g_3} -
4\alpha(\phi)\phi' \right] -e^2A^2R^2P^2 =0.
\end{equation}

It is impossible to find an analytical solution for eqs. (16-22). So, we will 
consider the solution of the cosmic string in the weak-field approximation. 
In fact, this approximation can only be justified if we consider the scalar 
field $\phi$ as a small perturbation on the gravitational field of the cosmic 
string. Thus, it may be expanded in terms of a small parameter $\epsilon$ 
about the values $\phi=\phi_0$ and $g_{\mu\nu}=\eta_{\mu\nu}$
\[
\phi=\phi_0 + \epsilon\phi_{(1)} 
\]
\[
g_{\mu\nu}= \eta_{\mu\nu} +\epsilon h_{\mu\nu} 
\]
\[
A(\phi)=A(\phi_0)+\epsilon A'(\phi_0)\phi_{(1)}
\]
\[
T^{\mu}_{\nu}=T^{\mu}_{(0)\nu}+\epsilon T^{\mu}_{(1)\nu}
\]
The term $(\phi')^2$ is neglected in the process of linearization of the 
Einstein eqs. (16-19). Moreover, in this approximation the $T^{\mu}_{(0)\nu}$ 
term is obtained from (20) by a limit process $\lambda \rightarrow \infty$ 
(see Linet \cite{li}). Therefore, it tends to the Dirac distribution on the 
surface $(t,z)$=constant. In this way, the linearized equations reduce to 
those of General Relativity \cite{vi}, except that in our case 
the terms $T^{\mu}_{(0)\nu}$ and $h_{\mu\nu}$ carry a conformal factor 
$A^2(\phi)$
\begin{equation}
\nabla^2 h_{\mu\nu} = 16\pi G (T^{\mu}_{(0)\nu} -\frac{1}{2}\eta_{\mu\nu}T_{(0)}),
\end{equation}
in the harmonic coordinates $(h^{\mu}_{\nu} - \frac{1}{2}\delta^{\mu}_{\nu}h)_{,\nu} =0$. 
Besides, equation (19) for the scalar field is also linearized 
\[
\nabla^2\phi_{(1)}=4\pi G\alpha(\phi_0)T_{(0)}.
\]
The solution for this equation is evident
\begin{equation} 
\phi_{(1)}=4G \mu A^2(\phi_0)\alpha(\phi_0)\ln(\rho/\rho_0).
\end{equation}
This solution matches with the vacuum solution iff
\begin{equation}
\kappa_{lin}=4G\mu \alpha(\phi_0)A^2(\phi_0).
\end{equation}
>From (14) we see that the only allowed values for $k_1$ are 
$k_1=0+ O(G^2\mu^2)$ and $k_1 = 4/3 + O(G^2\mu^2)$. The result which is physically meaningful is 
$k_1=0$ plus neglegible corrections of order $(G\mu)^2$; the other value 
corresponds to a non physical metric \cite{ga,vi}. 

The solution for eq. (23) differs from that found by Vilenkin \cite{vi} by a 
conformal factor $A^{2}(\phi)$. However, the procedure is the same as in his paper. After a coordinate transformation to bring the metric to the 
cylindrical system, we have  
\begin{eqnarray}
ds^2 & = & A^2(\phi_0)[ 1+8GA^2(\phi_0)\mu\alpha^2(\phi_0)\ln(\rho/\rho_0)]
[dt^2 - dz^2 -d\rho^2 \nonumber \\
& & -[1-8GA^2(\phi_0)\mu] \left( \frac{\rho}{\rho_0} \right)^2 d\varphi^2 ] ,
\end{eqnarray}
in which $\rho \geq 0$ and $0 \leq \varphi <2\pi$. 
Therefore, metric (27) represents an isolated scalar-tensor cosmic 
string so long as the weak-field approximation is valid. 

\subsection{Propagation of particles and light}

Concerning the light deflection in metric (27), it is easy to note that the 
angular separation of double image $\delta\varphi$ is given by 
$\delta\varphi = 8\pi GA^2(\phi_0)\mu = 8\pi \tilde{G}_0\mu$ (recall that 
$GA^2(\phi_0) = \tilde{G}_0$ is the effective Newtonian constant) and remains unchanged in the scalar-tensor gravity. For GUT strings, $\delta\varphi \sim 10^{-5}$ rad.  It means that, from the point of vue of this effect, it is 
impossible to distinguish a scalar-tensor cosmic string from a General 
Relativity one. 

The second remark is 
related to the fact that the  scalar-tensor cosmic string exerces a force 
on a non-relativistic test particle of mass $m$
\begin{equation}
f^{\rho} = - 4 mGA^2(\phi_0)\mu\alpha^2(\phi_0)\frac{1}{\rho} ,
\end{equation}
and it is always attractive. This force seems to be neglegible at present time 
(recall that $GA^2(\phi_0)\mu = \tilde{G}_0\mu \sim 10^{-6}$ for GUT strings,
and $\alpha^2(\phi_0) <10^{-3}$ \cite{da}). 

Let us consider the 
deflection of particles moving past the string. First of all, we  
rewrite metric (27) in terms of conformally Minkowskian coordinates
\begin{equation}
ds^2= (1+h_{00}) (dt^2 -dx^2-dy^2)
\end{equation}
where $h_{00}= 8GA^2(\phi_0)\mu \alpha^2(\phi_0)\ln [ (x^2 +y^2)^{1/2}]$ and, for simplicity, we consider $dz=0$. The factor $A^2(\phi_0)$ was absorbed by 
a rescaling.  
Metric (29) has a missing wedge  
of angular width $\Delta = 8\pi \tilde{G}_0 \mu$. The linearized geodesic equations in this metric have the form
\begin{eqnarray}
2\ddot{x} & = & -(1-\dot{x}^2-\dot{y}^2)\partial_x h_{00} \nonumber \\
 2\ddot{y} & = & -(1 -\dot{x}^2-\dot{y}^2)\partial_y h_{00}
\end{eqnarray}
where dots refer to derivative with respect to $t$. Now, if we choose the $x$-axis along which the particles are flowing with 
initial velocity $v_s$, the second of eqs. (30) can be integrated over the unperturbed trajectory $x=v_s t$ and $y=y_0$ . 
After the particles pass the string, their velocity gains a small 
$y$-component which can be computed in linear order in $\tilde{G}_0\mu$. For 
astrophysical applications, it may be interesting to express the relative 
velocity developed by the particles after  the string has passed between them. 
To transform to the frame in which the string has a velocity $v_s$ we make a 
Lorentz transformation \cite{vi2}. So, the final result is
\begin{equation}
u= 8\pi\tilde{G}_0\mu v_s\gamma + \frac{4\pi\tilde{G}_0\mu\alpha^2(\phi_0)}
{v_s\gamma}
\end{equation}
where $\gamma=(1-v_s^2)^{-1/2}$. The first term in (31) is equivalent to the 
expression of the relative velocity of particles flowing past a straight 
cosmic string in General Relativity \cite{ka} and it expresses the 
impulse velocity due to the conical deficit angle. The second term is due to 
the attractive force exerced by the scalar-tensor cosmic string.  
Let us compare the magnitudes of these two terms. String simulations show  
that $v_s \sim 0.15$ in the matter era \cite{al}. It means that the first term 
is about 45 times larger than the second one (recall that 
$\alpha^2(\phi_0) < 10^{-3}$ \cite{da}).  
At this point, it may be 
interesting to compare this effect in the scalar-tensor gravity with the relative velocity of particles in the background of a wiggly cosmic string in 
General Relativity \cite{va}. The fact that the wiggles produce an attractive 
force on the particles leads to similar correction to the usual expression 
\footnote{By ``usual" we mean the term produced by the unperturbed 
cosmic string in 
General Relativity.} for the relative velocity. 
However, one important difference is that in this case the term induced by 
the wiggles' force is ten times larger than the usual term 
in the matter era \cite{va}.  

\section*{Conclusions}

In this paper we studied the gravitational properties of a straight cosmic 
string in scalar-tensor gravities in the weak-field approximation. We 
showed that the propagation of particles and light in this background presents 
some analogy with the propagation of particles and light in the background of 
a wiggly cosmic string in General Relativity. But in the latter case the term 
of the relative velocity of particles moving past the string induced by the force produced by wiggles dominates over the usual term in the matter era. 
While in the former case the usual term dominates over the term produced 
by the force 
induced by the scalar-tensor gravity in the matter era. However, this 
situation may change in the early Universe because the field-dependent 
coupling strength between matter and the scalar field $\alpha^2(\phi)$ is 
expected to be of the same order of the coupling to matter of the tensor 
component of gravity. Finally, it may be worthwhile to point out that we are 
comparing the wiggly string in General Relativity to the scalar-tensor string 
at purely formal level. One should keep in mind that the formation of wakes in these 
two backgrounds has different physical origin: in the wiggly case, it is 
induced by the perturbations along the string (the wiggles) and in our case, 
it is induced by the scalar-tensor gravity.

\section*{Acknowledgements}

I am indebted to Prof. Bernard Linet for various enlightening discussions and 
for carefully reading this manuscript. I am also grateful to Dr. P. Teyssandier 
for valuable comments and suggestions. This work was supported by Conselho 
Nacional de Desenvolvimento Cient\'{\i}fico e Tecnol\'ogico (CNPq/Brasil).

\end{document}